\documentclass[11pt,twoside]{article}

\usepackage{asp2006}
\usepackage{epsf}
\usepackage{psfig}
\usepackage{lscape}

\begin{document}
\title{Estimation of the dust mass-loss rates from AGB stars in the Fornax and Sagittarius dwarf Spheroidal galaxies}   
\author{Eric Lagadec and Albert A. Zijlstra}   
\affil{University of Manchester, School of Physics and Astronomy}    
\author{M. Matsuura, P.A. Whitelock and J. Th. van Loon}   
\affil{NAOJ Tokyo, U Cape Town and U Keele}    
\begin{abstract} 
To study the effect of metallicity on the mass-loss of AGB stars, we have conducted mid-infrared 
photometric measurements of AGB stars in the Sagittarius and Fornax 
Dwarf Spheroidal Galaxies ([Fe/H]=-1.1 and -1.3) with the 10-micron camera
of VISIR at the VLT.
These observations combined with previous near-infrared photometric 
measurements allow us to estimate mass-loss rates in these galaxies. We show here that the observed AGB display dust-driven mass-loss. Dust mass-loss rate  are found to be in the range 0.2$\times10^{-10}$-1.3$\times10^{-8}$ M$_{\odot}$yr$^{-1}$ for the observed AGB stars in SgrD and around 5$\times10^{-11}$ M$_{\odot}$yr$^{-1}$ for the observed star in Fornax.
\end{abstract}
\vspace{-1.2cm}
\section{Introduction}
During the late stages of their evolution, stars with initial mass on the main sequence in the range 0.8-8 M$_{\odot}$ reach the Asymptotic Giant Branch phase. This phase of stellar evolution is characterised by a strong mass-loss, the star expelling dust and gas at rates up to $10^{-4}M_{\odot}$yr$^{-1}$. This is really important for the chemical evolution of our Galaxy. It has been shown that mass-loss from this stars contributes to around half of the gas recycled by stars (Maeder 1992).

 The mass-loss mechanism of AGB stars is a two step process. Shocks due to pulsations from the star produce gas surdensities that can trigger the formation of dust. Then radiation pressure on dust make this dust escape the gravity of the stars. Gas is thus also expelled thanks to friction with dust. 
The importance of AGB stars on the chemical evolution of other galaxies is poorly known. In particular, the effect of metallicity on the mass-loss rates of AGB stars is not well known. Indeed, in low metallicity environments, less dust is expected to form. Mass-loss rates should thus be lower in low metallicity environments. A theoritical work by Bowen and Willson (1991) predicts that for metallicities below 0.1 Z$_{\odot}$ dust-driven winds should be negligeable.

Observations in the Magellanic Clouds and the Galaxy have shown that the total mass-loss rates (dust+gas) were similar in the three galaxies (van Loon 1999). In fact the dust mass-loss rates are smaller in the Magellanic Clouds, but the gas-to-dust mass ratio are higher in these galaxies with low metallicities (0.3 Z$_{\odot}$ and 0.5 Z$_{\odot}$ for the Small and Large Magellanic Clouds respectively). In order to obtain better constraints on the effect of metallicity on the mass-loss rates from AGB stars, and to know if dust-driven mass loss can occur at very low metallicities, we need to study the mass-loss rates from AGB stars with low metallicities. 

We thus observed AGB stars in the Fornax and Sagittarius Spheroidal dwarf (hereinafter SgrD) galaxies. Those galaxies have low metallicities, 0.008 Z$_{\odot}$ and 0.3 Z$_{\odot}$ respectively. Furthermore the distance of these galaxies is quite known, making the estimation of the mass-loss rates easier than in the Galaxy.

\section{Observations and target selection}
The observations were made with the mid--infrared imager VISIR (Lagage et al. 2004), at the VLT (Paranal, Chile). These observations were carried in visitor mode during 4 nights in July 2005. We observed 28 AGB stars in SgrD and 2 in Fornax. Due to bad weather and the weakness of the signal for some of the observed stars, we will present here results for 9 stars in SgrD and 1 in Fornax.

To reduce the background emission from the sky and the telescope, we used the standart mid-infrared technique of chopping and nodding. We observed all the stars with the Visir PAH1 filter ($\lambda_c$=8.59$\mu$m, $\Delta$$\lambda$=0.42$\mu$m). The brightest stars were also observed with the Visir PAH2 filter  ($\lambda_c$=11.25$\mu$m, $\Delta$$\lambda$=0.59$\mu$m).

\begin{table*}
\caption[]{\label{log_obs} Observed SgrD and Fornax targets:
names, adopted coordinates,  and  photometry.
JHK is taken from 2MASS.}
\begin{flushleft}
\begin{tabular}{llllllllllllllll}
\hline
Adopted name & RA & Dec  
& J & H & Ks  \\
  &\multicolumn{2}{c}{(J2000)} & mag & mag & mag  \\
\hline
2mass01        &18 40 57.5  & $-$27 34 22.7&12.479 &10.774  &9.308 && & \\
2mass15        &18 46 51.6  & $-$28 45 48.9&13.060 &11.123  &9.599 & & & \\
SgrCn28        &18 51 11.08 & $-$31 21 56.3&13.062 &11.528  &10.32 & & & \\
SgrCn14        &18 51 41.05 & $-$30 03 37.7&12.090 &10.570  &9.42  & \\
2mass50        &19 04 35.62 & $-$31 12 56.4&14.227 &11.956  &10.168 & & \\
2mass63        &19 10 39.87 & $-$32 28 37.3&11.957 &9.991   &8.349  & & \\
2massSGC21     &20 20 27.66 & $-$14 49 27.1&11.850 &10.143  &8.707  & & \\
SgrCn16        &18 53 40.98 & $-$29 34 22.9& 13.258&11.599  &10.182& & & \\
WMIF-C18       &19 09 39.03 & $-$29 56 56.1&11.188 &9.946   &9.201  & & \\

 & & & &  & & & & \\
02385056 &02 38 50.56 & $-$34 40 31.9&16.106 &14.525  &12.879 & & & \\

\hline \\
\end{tabular}
\end{flushleft}
\end{table*}

\vspace{-1.3cm}
\section{Methods to determine mass-loss rates from infrared colors}
\label{other_methods}
Mass-loss rates from AGB stars in the Galaxy have already been estimated using mid-infrared photometry. From a survey of Miras in the South Galactic Cap, Whitelock et al. (1994) showed that the mass-loss rates from these stars and the K-[12] colour were tightly correlated (see Fig.21 in their paper), where [12] is the IRAS 12$\mu$m magnitude. This can be explained by the fact that the K magnitude is correlated with the emission from the star and the [12] magnitude with emission from dut in the circumstellar envelope. The K-[12] colour is thus a measurement of the optical depth of the envelope, that is increasing with mass-loss rate.

Mass-loss rates from Galactic AGB stars have also been estimated using dust radiative transfer models (e.g. Le Bertre, 1997). 
The optical depth is a free parameter in this models and is proportional to the mass-loss rate. Thus fitting an observed SED using a radiative transfer model give an estimate of the optical depth and thus of the mass-loss rate. Such models of AGB stars in the Galaxy have shown that the derived values of the mass-loss rates were compatible with values obtained with different method (e.g. CO observations).

\section{Our methods}
\begin{figure}
\plottwo{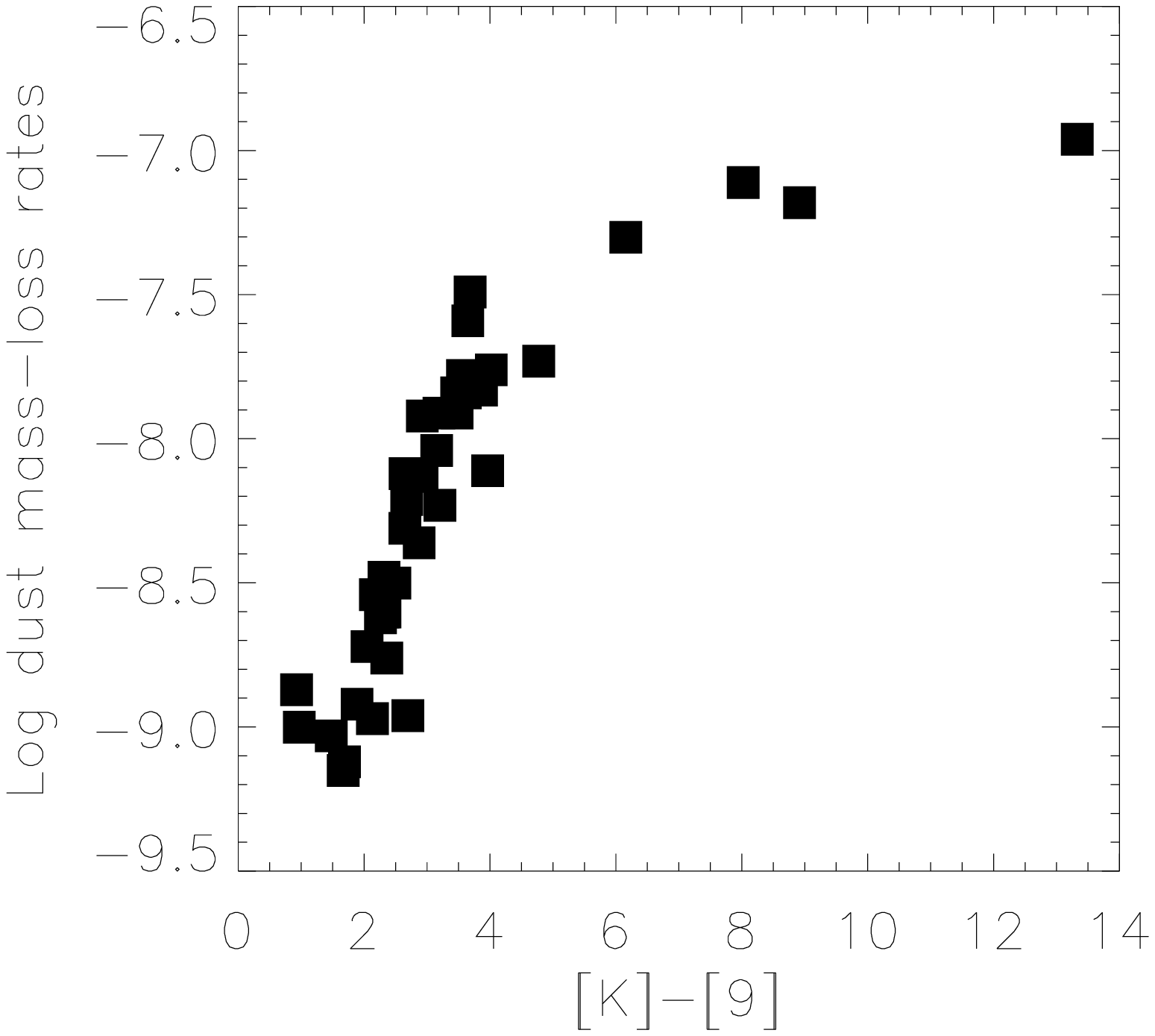}{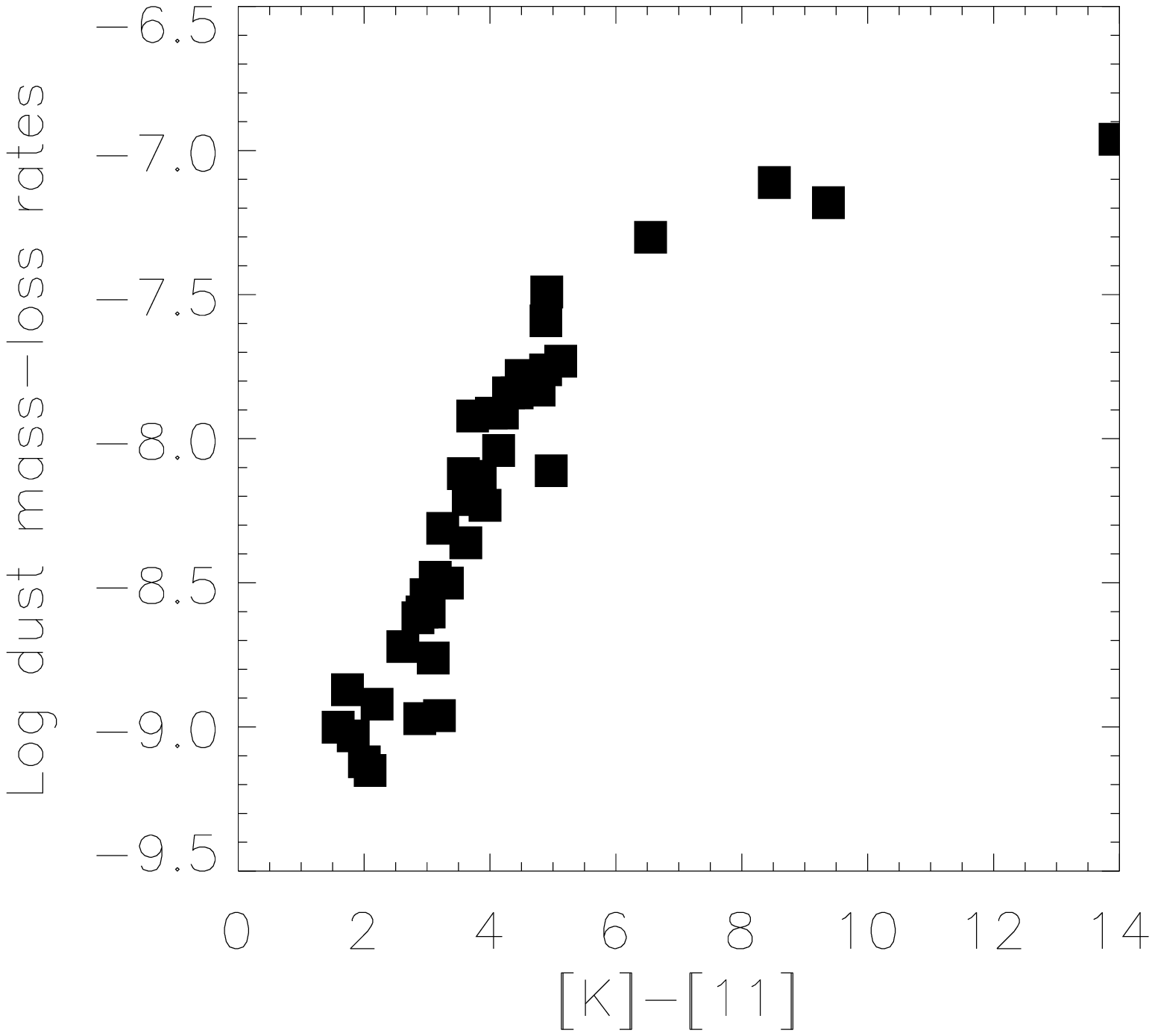}
\caption{\label{massloss_whitelock_9} Dust mass-loss rates as a function of K-[9[ colours of stars from the sample in Whitelock et al.(1994).}  
\end{figure}

As mentioned in section \ref{other_methods}, it has been showed that the K-12 colour was tightly correlated with the mass-loss rates for Galactic stars (Whitelock et al. 1994). In their work, the 12$\mu$m magnitude is determined using IRAS measurement. Zijlstra et al. (1996) have shown that this relation can be quantified as:
\begin{equation}
\label{mass-loss}
log(\dot{M}_{dust})=0.57\times (K-[12])-10.35
\end{equation}
The sources we observed are faint and haven't been observed by IRAS. We carried out our observations using the filters with wavelength close to 12 $\mu$m having the best sensitivity. The two filters we selected have wavelengths centered at 8.59 and 11.25$\mu$m respectively. To have an equivalent relation  to Eq \ref{mass-loss} using these filters, we retrieved all the avalaible IRAS spectra used by Whitelock et al. We then convolved these spectra with the VISIR transmission curves of the filters used. We thus get an estimation of the [9] and [11] colours of the sources observed in the South Galactic Cap. It enables us to confirm there is a tight correlation between the K-[9] and K-[11]  (Fig.\ref{massloss_whitelock_9}) and dust mass-loss rates. Note that Whitelock et al. (1994) used total mass-loss rates (dust+gas), using a gas to dust mass ratio of 200. As what we observed at 9 and 11$\mu$m is mostly due to emission from dust, what we determine is dust mass-loss ratios. The gas to dust mass ratios in the envelope of AGB stars in the observed galaxies being poorly known, we prefer to talk about dust mass-loss rates.
We thus get:

\begin{equation}
\label{mass-loss}
log(\dot{M}_{dust})=0.65\times (K-[9])-11.42
\end{equation}
and :
\begin{equation}
\label{mass-loss}
log(\dot{M}_{dust})=0.52\times (K-[11])-11.33
\end{equation}

To estimate the mass loss rates of the stars we observed, we also computed some dust radiative transfer models using the DUSTY 1D code (Ivezic et al., 1999).For every models, the density distribution was taken as the one of a radiation-driven wind. The other free parameter were chosen to be similar to typical AGB stars (T$_{eff}$=2800K, a=0.1$\mu$m, mixture of amorphous carbon (95\%) and SiC (5\%)). 
 To estimate the dust mass-loss rates, we assumed that the expansion velocity of the dust shell was set to 5 km.s$^{-1}$. This value is smaller than what is usually observed in the Galaxy as low metallicity AGB stars have lower expansion velocities.

The optical depth of the envelope is proportional to the dust mass-loss rate. We can thus associate a mass-los rate with a model of a given optical depth so that for our models:
\begin{equation}
\label{mass1}
\dot{M}_{dust}=1.07 \times 10^{-9}(K-[11])-1.8 \times 10^{-9}(K-[11])^2+9.5 \times 10^{-10}(K-[11])^3
\end{equation}
\vspace{-0.7cm}
\begin{equation}
\label{mass2}
\dot{M}_{dust}=6.2 \times 10^{-10}(K-[9])-1.7 \times 10^{-9}(K-[9])^2+1.2 \times 10^{-9}(K-[9])^3
\end{equation}
We thus calculated a benchmark of DUSTY models with this input parameters and different optical depths.
The outputs of these models are spectra which we convolved with the 2MASS K$_s$ and VISIR PAH1 and PAH2 filters to determine the K-[9] and K-[11] colours associated with the different models. These colours then directly give us an estimation of the mass-loss rates using relation (\ref{mass1}) and (\ref{mass2}).
%
\section{Dust mass-loss rates}
 The derive values for the dust mass-loss rates are given in Table \ref{mass_loss}
The first main result of this study is that the observed stars in the Sgr D and Fornax are reddened , indicating the presence of a dusty envelope, and thus that mass-loss occurs in these stars.
. The estimated dust mass-loss rates are found to be in the range 0.2$\times10^{-10}$-1.3$\times10^{-8}$ M$_{\odot}$yr$^{-1}$ for the observed AGB stars in SgrD and around 5$\times10^{-11}$ M$_{\odot}$yr$^{-1}$ for the observed star in Fornax. The values obtained with the two different methods are of the same order or magnitude.

\begin{table*}[h!!!!]
\caption[]{\label{mass_loss} Mass-loss rates. PAH1 and PAH2 are the flux observed in Jy.$\dot{M}_1$ and $\dot{M}_2$ are the mass-loss rates obtained using the first method using the K-[9] and K-[11] colours respectively. $\dot{M}_3$ and $\dot{M}_4$ are those obtained using the radiative transfer model ( mass-loss rates in units of $\times10^{-9}$ M$_{sun}$yr$^{-1}$).}
\begin{flushleft}
\begin{tabular}{llllllllllllllll}
\hline
Adopted name & PAH1 & PAH2 & K-[9]& K-[11] & $\dot{M}_1$& $\dot{M}_2$ &$\dot{M}_3$ &$\dot{M}_4$   \\

\hline
2mass01        &0.08  &0.01    &2.3 &   &2.8&      &7.0& \\
2mass15        &0.02  &        &1.1 &   &0.7&      &0.2&     \\
SgrCn28        &      & 0.01   &    &1.6&   &0.6   &   &1.0 \\
SgrCn14        & 0.060& 0.052  &2.1 &2.5&2.2&1.7   &4.9& 6.3\\
2mass50        & 0.018&        &1.6 &   &1.3&      &1.6&     \\
2mass63        & 0.177& 0.225  &2.2 &3.0&2.5&2.8   &5.9&12.7\\
2massSGC21     & 0.089& 0.063  &1.8 &2.0&1.6&1.0   &2.6& 2.6\\
SgrCn16        & 0.018& 0.01   &1.6 &   &1.3&      &1.6&    \\
WMIF-C18       & 0.009& 0.0104 &2.3 &   &2.8&      &7.0& \\

 & & & &  & & & & \\
02385056       &0.0008&        &0.9 &   &0.02& &0.06\\

\hline \\
\end{tabular}
\end{flushleft}
\end{table*}
\section{Discussion and conclusion}
In  this work we have shown that AGB stars in the Sgr D Sph and Fornax galaxies were losing mass. Dust-driven mass-loss is thus observed for this AGB stars in low metallicity galaxies. We estimate dust mass-loss rate in the range 0.2$\times10^{-10}$-1.3$\times10^{-8}$ M$_{\odot}$yr$^{-1}$ for the observed AGB stars in SgrD and around 5$\times10^{-11}$ M$_{\odot}$yr$^{-1}$ for the observed star in Fornax.

To convert this value to total (gas+dust) mass-loss rates, one needs to know the gas-to-dust mass ratio. This ratio is not well known for Galactic stars, so that i.e. for IRC\,+10216, the best studied AGB stars, values in the range 170-700 are found in the literature (Mensh'chikov et al., 2001). It is difficult to determine this ratio in other galaxies. However, van Loon (2000) as shown that this ratio was linearly decreasing with metallicity. A simplified view to explain this is that the fraction of metals avalaible for the formation of dust is proportional to metallicity, so that at lower metallicity, less dust is formed. If we consider that the metallicity in Sgr D Sph and Fornax are 0.3 and 0.8 Z$_{\odot}$ respectively, then we can crudely estimate that the dust mass-loss rate in these galaxies are 1000 and 3000 times higher than in the Galaxy. Converting  the dust mass-loss rates we estimated would thus lead to  total mass-loss rates of the order of $10^{-5}$- $10^{-4}$M$_{\odot}$yr$^{-1}$. These rates are similar to the highest in the Galaxy, but one has to consider these total mass-loss rates as really crude estimates. For example, the drift velocity (the velocity difference between the gas and the dust fluid) is not taken into account in these estimations. This drift velocity is certainly smaller at low metallicity so that the values estimated just above are certainly overestimated. To better understand the mass-loss of AGB stars in these galaxies, we thus need to improve our models and study a larger sample of AGB stars. Such observations have been carried out or are schduled with Spitzer or at the VLT and will enable to get constraints on the gas-to-dust mass ratio of AGB stars in these galaxies and to study a wider range of infrared colours (and thus to study the evolution of the mass-loss on the AGB). This will be presented in forthcoming works.

\end{document}